\documentclass[aps,pre,notitlepage,preprint,11pt,a4paper]{revtex4-1}

\usepackage{graphicx}
\usepackage{amsmath}
\usepackage{amssymb}
\usepackage{color}
\usepackage{upgreek}
\usepackage{mcode}
\usepackage[]{natbib}
\bibpunct{(}{)}{;}{a}{,}{,}

\newcommand{\dd}{\ensuremath{\mathrm{d}}}
\newcommand{\dbyd}[2]{\ensuremath{\frac{\dd #1}{\dd #2}}}

\begin{document}

\title{A Cell-Level Mechanism of Contrast Gain Control}
\author{Linus J. Schumacher}
\affiliation{Wolfson Centre for Mathematical Biology, Mathematical Institute, University of Oxford, Radcliffe Observatory Quarter, Woodstock Road, Oxford, OX2 6GG, United Kingdom}
\affiliation{Computational Biology Group, Department of Computer Science, University of Oxford, Wolfson Building, Oxford, OX1 3QD, United Kingdom}
\author{Geoff K. Nicholls}
\affiliation{Department of Statistics, University of Oxford, 1 South Parks Road, Oxford, OX1 3TG, United Kingdom}
\date{\today}

\begin{abstract}
		 The gain of neurons' responses in the auditory cortex is sensitive to contrast changes in the stimulus within a spectrotemporal range similar to their receptive fields~\citep{Rabinowitz2012}, which can be interpreted to represent the tuning of the input to a neuron. This indicates a local mechanism of contrast gain control, which we explore with a minimal mechanistic model here.
		Gain control through noisy input has been observed \textsl{in vitro}~\citep{Chance2002} and in a range of computational models~\citep{Ayaz2009,Longtin2002}. We investigate the behaviour of the simplest of such models to showcase gain control, a stochastic leaky integrate-and-fire (sLIF) neuron, which exhibits gain control through divisive normalisation of the input both with and without accompanying subtractive shift of the input-response curve, depending on whether input noise is proportional to or independent of its mean. To get a more direct understanding of how the input statistics change the response, we construct an analytic approximation to the firing rate of a sLIF neuron constituted of the expression for the deterministic case and a weighted average over the derived approximate steady-state distribution of conductance due to poissonian synaptic inputs.
		This analytic approximation qualitatively produces the same behaviour as simulations and could be extended by spectrotemporally tuned inputs to give a simple, physiological and local mechanism of contrast gain control in auditory sensing, building on recent experimental work that has hitherto only been described by phenomenological models~\citep{Rabinowitz2012}. By comparing our weighted average firing rate curve with the commonly used sigmoidal input-response function, we demonstrate a nearly linear relationship between both the horizontal shift (or stimulus inflection point) and the inverse gain of the sigmoid and statistics derived from the sLIF model parameters, thus providing a structural constraint on the sigmoid parameter choice.
\end{abstract}

\maketitle


\section{Introduction -- Contrast Gain Control}
	\begin{figure}[b]
	  \includegraphics[scale=0.35]{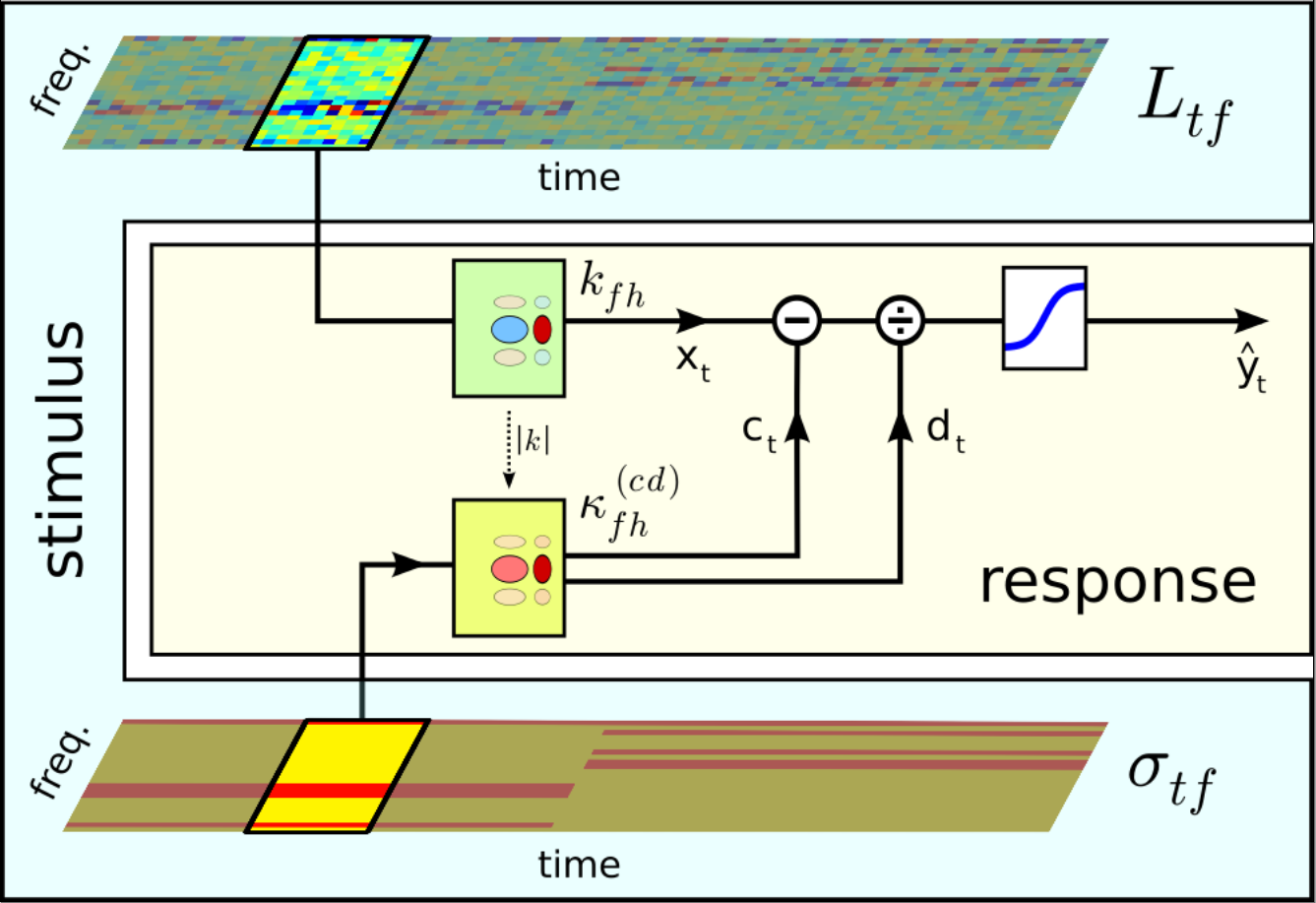}
	  \caption{\textbf{Spectrotemporal contrast kernel model by \citet{Rabinowitz2012}:} The stimulus spectrogram $L_{tf}$ is convolved with a linear spectrotemporal kernel $k_{fh}$ (equivalent to the STRF) and passed through sigmoidal response function to give the predicted firing rate $\hat y_t$. The horizontal shift, or stimulus inflection point, $c$ and the inverse gain $d$ of the sigmoid are functions of stimulus contrast $\sigma_{tf}$, and their relationship is described by a single contrast kernel $\kappa_{fg}^{(cd)} \approx |k_{fh}|$. Reproduced with permission.}
	  \label{figGainKernelModel}
	\end{figure}

	Animals can process sensory information over a wide range of stimulus intensity contrasts. To do so, the gain of their neural response adjusts itself to variations in spatiotemporal (in vision) or spectrotemporal contrast (in hearing). One such mechanism of gain control, divisive normalisation, has been successfully applied in models of the visual system for a long time, but been studied much less in the auditory system. Recently it has been shown~\citep{Rabinowitz2011a,Rabinowitz2012} that a large part of gain control of neurons in the auditory cortex can be accounted for by horizontally shifting and divisively normalising the neural response, i.e. firing rate, in proportion to the stimulus contrast (Fig.~\ref{figGainKernelModel}). The exact relationship can be described by ``gain contrast kernels''~\citep{Rabinowitz2011a}. Furthermore, it was found that the gain control is most sensitive to contrast in spectrotemporal regions where the neuron's response is also high in magnitude~\citep{Rabinowitz2012,Rabinowitz2011a}. In other words, the contrast gain control seems to operate mainly within the spectrotemporal receptive field (STRF) and be insensitive to variations in stimulus contrast outside the STRF. This phenomenological description of gain control is indicative of local computation and raises the possibility that the underlying physiological or biophysical mechanism operates on a single cell rather than on a population level.
	
	Gain control has also been observed in mathematical and computational neuron models, the simplest relevant one being the leaky integrate-and-fire (LIF) neuron with noisy synaptic input. This model springs to mind as a good candidate for a local mechanism of contrast gain control. Starting from a LIF model (as implemented in, for example,~\citep{Ayaz2009}, but without constructing any pools of normalising or modulatory neurons) we examine gain control in response to change in input statistics, aiming to find analytical approximations to the simulation results wherever possible in order to get a better handle on the model. By identifying qualitative and quantitative similarities between the stochastic LIF model and the gain contrast kernel model, we suggest their equivalence. The stochastic LIF model is presented as a hypothesis for the mechanism that could underlie the gain contrast kernels
	
\section{Methods}

	\subsection{Leaky integrate and fire (LIF) neurons} 
		We start with a basic neuron model that is analytically tractable: The leaky integrate-and-fire neuron model is generally regarded as the simplest description of a nerve cell that produces any realistic output. The membrane potential $V$ follows the differential equation
		\[C\dbyd{V(t)}{t} = I(t) - gV(t)\]
		where $C$ is the membrane capacitance, $g$ the membrane conductance and $I(t)$ the input current, which could be current flowing across the membrane through ion channels or current injected in a patch-clamp experiment. When the input current is large enough, the membrane potential rises until it reaches a threshold $V_{th}$, upon which a spike is recorded and the potential is reset to $V_{reset} = 0$, representing the firing of an action potential. For a constant input current, the firing rate (number of spikes per time) is given by the expression~\citep{Koch1998}
		\begin{equation}
		f(I) = \begin{cases} 
				0,  & I \le gV_{th} \\
	 			 {[} t_{ref} - \frac{C}{g}\log(1-\tfrac{gV_{th}}{I}) {]}^{-1}, & I > gV_{th} 
			\end{cases} \label{eqn_f(I)}
		\end{equation}
		where $t_{ref}$ is the refractory period, i.e., the length of time spiking is suppressed after an action potential has been fired. Note that the model as such does not exhibit a refractory period, it has to be added to achieve saturation of the firing rate. One could set $t_{ref}$ to a physiologically realistic value on the order of $1~\mathrm{ms}$, but for the purpose of comparing to simulations we simply set it to the simulation time step $\Delta t$.
		
	\subsection{stochastic LIF (sLIF) neurons}\label{sLIFmodel}
		Having a firing rate expression for the deterministic LIF neuron, we now want to add noise to the input. An extension to the LIF model that makes it more physiological is to let the membrane conductance $g$ vary with stochastic synaptic input, which can be excitatory or inhibitory. To compare with simulations, we adapt the implementation from \citet{Ayaz2009} in this section, and stay close to their notation throughout this paper. Whenever a spike arrives at an excitatory (inhibitory) synapse, e.g. through a Poisson process with rate $\lambda_e$ ($\lambda_i$), the conductance increases instantaneously by a fixed absolute amount $\Delta g_e$ ($\Delta g_i$), representing the assumption that a fixed amount of neurotransmitter is released and opens a fixed amount of postsynaptic ion channels. Between spikes, the excitatory (inhibitory) part of the conductance decays to zero with time constant $\tau_{g_e}$ ($\tau_{g_i}$). The synaptic conductance thus evolves independently of the membrane potential and any reversal or threshold values thereof. However, the conductance terms are now no longer constant, so the differential equation describing the evolution of the membrane potential becomes
		\[C\dbyd{V(t)}{t} = I_{FF}(t) + g_L[V_L - V(t)] + g_e(t)[E_e - V(t)]  + g_i(t)[E_i - V(t)]\]
		where $E_e$ ($E_i$) is the reversal potential of the excitatory (inhibitory) ion channels, $I_{FF}$ is the feed-forward current and $V_L$ ($=0$ earlier) is the resting membrane potential, which is also assumed to be the value of $V_{reset}$.
		
		If we shift the potential by $V_L$ and separate the voltage dependent and independent terms, we get back our original equation
		\[C\dbyd{V(t)}{t} = I(t) - g(t)V(t)\]
		only that now we have `effective' current and conductance:
		\begin{align}
			g(t) &= g_L + g_e(t) + g_i(t) \label{eqn_g}\\
			I(t) &= I_{FF}(t) + g_e(t)[E_e - V_L] - g_i(t)[V_L - E_i] \label{eqn_I}
		\end{align}
		Ideally we would now directly derive a stochastic version of the deterministic firing rate equation (\ref{eqn_f(I)}), but this is intractable. Instead, we will try and approximate the distribution of $g_e(t)$ and $g_i(t)$. From this, we might be able to work out the distributions of $g(t)$ and $I(t)$, and so on, hopefully arriving at a distribution of $f(I)$ -- though a simpler approach will prove far more tractable, as we will see in the rest of this section.
		
		The model complexity is deliberately kept low for the sake of tractability and suffices for the purpose of our argument. However, more complex neuron models, such as the Morris-Lecar have been shown to be equivalent to a stochastic LIF neuron \citep{Ditlevsen2012}, giving additional justification to our model choice.
		
		\subsubsection{Mean-field failure}
			 Na\"ively one might try a mean-field approach and substitute the steady-state values $\langle g_{e,i}\rangle = \Delta g_{e,i} \lambda_{e,i} \tau_{g_{e,i}}$ (assuming synaptic input is Poisson distributed with rate $\lambda$) into (\ref{eqn_f(I)}), but in practice this will often give zero firing rate even when simulations show spiking. This is because even when the condition $\langle I \rangle > \langle g \rangle V_{th}$ is not satisfied, fluctations, which are not captured by the mean field approach, can drive the membrane potential above threshold.
			 
		\subsubsection{Analytical approaches in the literature}
			Decades of work have been dedicated to obtain analytical expressions for the firing rate of an sLIF neuron~\citep{Koch1998}. Approaches include use of the Fokker-Planck equation (\citep{Ermentrout2010}, section 10.2, and references therein), first order corrections to the first passage time~\citep{Brunel1998}, the Volterra integral equation~\citep{Buonocore2010} and coupling of excitatory and inhibitory noise~\citep{Lanska1994,Lansky1995,Longtin2002}. And while these are all valuable achievements in one way or another, they typically involve integrals that can only be evaluated numerically (such as error functions~\citep{Yu2003,Ostojic2011}), infinite sums, gamma functions or other mathematical unpleasantries. These may allow the neuroscientist to do away with running exhaustive simulations (at least for those special cases that these approaches have been tailored to), but they generally don't provide a good insight into how a change in statistics of the noisy input affect the output.

			For an Ornstein-Uhlenbeck (OU) neuron a linear approximation for the firing rate in terms of input mean and variance can be found \citep{Lansky2001,Yu2003,Sacerdote2013}. However, in the implementation of a stochastic LIF neuron here and elsewhere~\citep{Ayaz2009}, the membrane Voltage is not described by an OU process. Rather, it is the membrane conductance that can be described by an OU process, as we will show in Section~\ref{distributionDerivations}. The membrane voltage differential equation is thus coupled to the solution of this OU process. If one assumes the conductance becomes a random variable itself, we suspect one could arrive at a Chan-Karolyi-Longstaff-Sanders (CKLS), or Brennan-Schwartz (BS), process~\citep{Chan1992,Brennan1980}, in which voltage fluctuations are dependent on the input statistics and the voltage itself. To our knowledge, the mathematical behaviour of such a CKLS/BS neuron has never been studied.  
\clearpage
		\subsubsection{Parameter values used} \label{modelParameters} When simulating the sLIF model to test our analytical approximations, we used the following parameter values, as in~\citep{Ayaz2009}.
			\begin{center}
				\begin{tabular}{l  l  r  l}
					\textbf{Parameter} & \textbf{Description} & \textbf{Value} & \textbf{Units} \\ \hline 
					$C$ & membrane capacitance & 740 & pF \\ 	
					$g_L$ & resting (or leak) conductance & 20 & nS \\ 
					$V_L$ & resting (and reset) membrane potential & -70 & mV \\ 
					$V_{th}$ & threshold potential for firing an action potential & -52 & mV \\ 
					$E_e$ & reversal potential of excitatory conductance & 0 & mV \\ 
					$E_i$ & reversal potential of inhibitory conductance & -80 & mV \\ 
					$\Delta g_e$ & instantaneous change in excitatory conductance & 0.16$g_L$ & $[g_L]$ \\ 		
					$\Delta g_i$ & instantaneous change in inhibitory conductance & 0.48$g_L$ & $[g_L]$ \\ 
					$\tau_{g_{e,i}}$ & time constant of synaptic conductances & 5 & ms \\ 	
					$\Delta t$ & time step & 0.05 & ms \\ 
					$T$ & typical length of each simulation of neural activity & 1 & s \\ 
				\end{tabular}
			\end{center}
			
	\subsection{Approximations of the steady-state distribution of stochastic synaptic input} \label{distributionDerivations}
		To understand the influence of fluctuations in synaptic input on the firing rate of our sLIF neuron, consider the distribution of conductances, excitatory or inhibitory, here generically denoted by $g$. In an actual implementation of the sLIF model as described in section \ref{sLIFmodel}, time will be discretised and the evolution of each of the excitatory and inhibitory conductance terms can be described by an AR(1)-process of the form
		\[g_t = c + \phi g_{t-\Delta t} + \sigma_\xi\xi_t\]
		where $c$ is a constant, $\phi$ is the autoregressive parameter and $\xi_t$ is white noise with variance $\sigma_\xi$. This is based on the assumption that poissonian synaptic input can be approximated as 
		 \[\Delta g \cdot \mathcal{P} (\lambda\Delta t) \approx \Delta g \cdot \mathcal{N}(\lambda\Delta t,\sqrt{\lambda\Delta t}) = c + \sigma_\xi\xi_t\] 
		 where $\mathcal{P}$ and $\mathcal{N}$ are Poisson and normally distributed random variables, respectively (Note we use $\mathcal{P}$ and $\mathcal{N}$ to denote both the random variable and its distribution). So $c = \sigma_\xi^2 = (\Delta g)^2\lambda \Delta t$ and $\phi = e^{-\Delta t/\tau_g} \approx 1 - \Delta t/\tau_g$ for time steps $\Delta t << \tau_g$. Substituting these values into the expressions for the mean $\mu = c/(1-\phi)$ and variance $\sigma^2 = \sigma_\xi^2/(1-\phi^2)$ of an AR(1) process and expanding to first order in $\Delta t/\tau_g$ we see that steady-state value of each of the conductance terms is approximately normally distributed with mean and variance 
		\begin{align}
			\mu_g &= \Delta g \lambda \tau_g \label{eqn_mu_g}\\
			\sigma_g^2 &= (\Delta g)^2\lambda\tau_g/2 \label{eqn_sigma_g}
		\end{align}
		One can also write the evolution of $g$ in the continuous time limit as an Ornstein-Uhlenbeck (OU) process. Starting from our discrete time AR(1)-process 
		\[g_t = (1-\Delta t/\tau_g) g_{t-\Delta t} + \Delta g \lambda \Delta t + \Delta g\sqrt{\lambda\Delta t}\xi_t\] 
		and taking the limit $\Delta t \rightarrow 0$ we get the expression 
		\[\dd g = \Phi(\mu_g - g)\dd t + \sigma_W \dd W \] 
		where $\Phi=1/\tau_g$, the mean $\mu_g$ is as in \eqref{eqn_mu_g} and $W$ is a Wiener process with variance $\sigma_W^2 = \Delta g^2 \lambda$ . The variance around the steady-state value in an OU-process is $\sigma_g^2 = \sigma_W^2/2\Phi$, which gives us the same value for $\sigma_g$ as in (\ref{eqn_sigma_g}).
		
		We make the steady-state assumption here, but LIF neurons with dynamic inputs have been shown to have similar gain control~\citep{Ly2009}.
			
	\subsection{Weighted average of deterministic LIF neuron firing rate}
		We derived in section \ref{distributionDerivations} that the excitatory and inhibitory conductances are approximately normally distributed over time. By ergodicity, or by wide-sense stationarity, the distribution of a conductance value over time is equal to distribution of conductance values at a particular time point. Hence, we can estimate the firing rate of a stochastic LIF neuron by taking the average of the firing rate expression for the deterministic case (\ref{eqn_f(I)}) over the joint probability distribution of $g_e$ and $g_i$ (assumed independent), and recall that we shifted all potential terms by $V_L$ in the sLIF model construction, i.e.
		\begin{equation}
			f(I) \approx \langle{[} t_{ref} - \frac{C}{g}\log(1-\tfrac{g(V_{th} - V_L)}{I}) {]}^{-1}\rangle_{\mathcal{N}(\mu_{g_e},\sigma_{g_e}) \cdot \mathcal{N}(\mu_{g_i},\sigma_{g_i})}, I > g(V_{th} - V_L)
			\label{eqn_weightedAverage}
		\end{equation}
		where $g$ and $I$ are given by equations (\ref{eqn_g}) and (\ref{eqn_I}), respectively, and $\mu_{g_{e,i}}, \sigma_{g_{e,i}}$ are given by (\ref{eqn_mu_g},\ref{eqn_sigma_g}).

\section{Results}
	\begin{figure}[b] 
	 \includegraphics[scale=0.75]{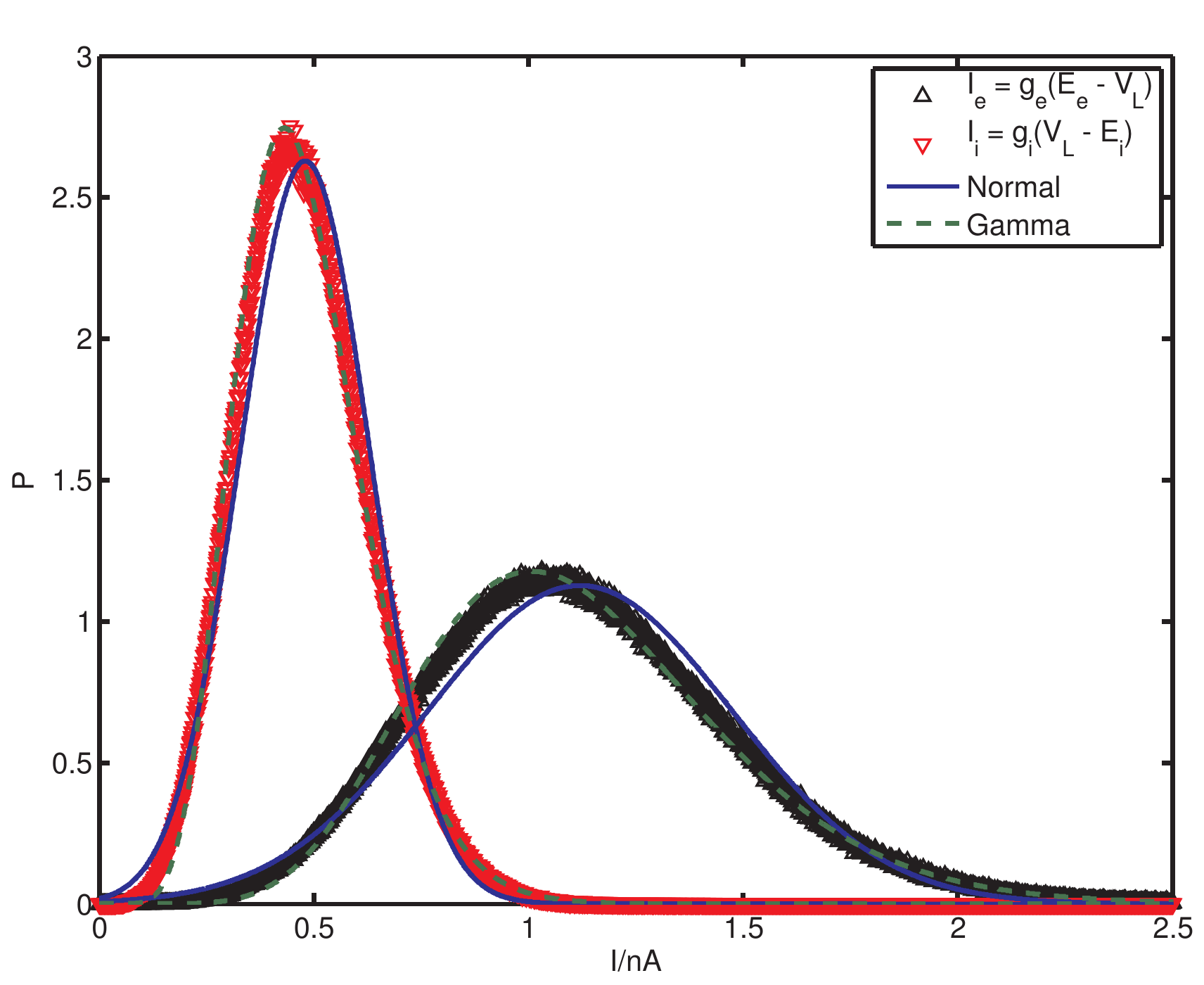}
	 \caption{\textbf{Steady-state distributions  of the excitatory ($g_e$) and inhibitory ($g_i$) conductances in a sLIF neuron model.} Excitatory ($I_e$) and inhibitory ($I_i$) synaptic currents are approximated by a normal and a gamma distribution. The model differential equation, in terms of the parameters plotted, is $C\dbyd{V}{t} = d - \frac{n}{V_{th} - V_L}V$ and the firing rate for the deterministic case is $f(I) = {[} t_{ref} - \frac{C(V_{th} - V_L)}{n(g)}\log(1 - r(I,g)) {]}^{-1}$. Distribution parameters are \textsl{not} fitted, but derived as described in the main text (\ref{eqn_mu_g},\ref{eqn_sigma_g}). Data shown for synaptic input rates $\lambda_e = \lambda_i = 1~\mathrm{kHz}$, but approximations were qualitatively similar for rates as low as $0.5~\mathrm{kHz}$, and better the higher the rates.
	 			}
	 \label{figConductanceDistributions}
	\end{figure}
	
	\begin{figure}[]
	 \includegraphics[scale=0.75]{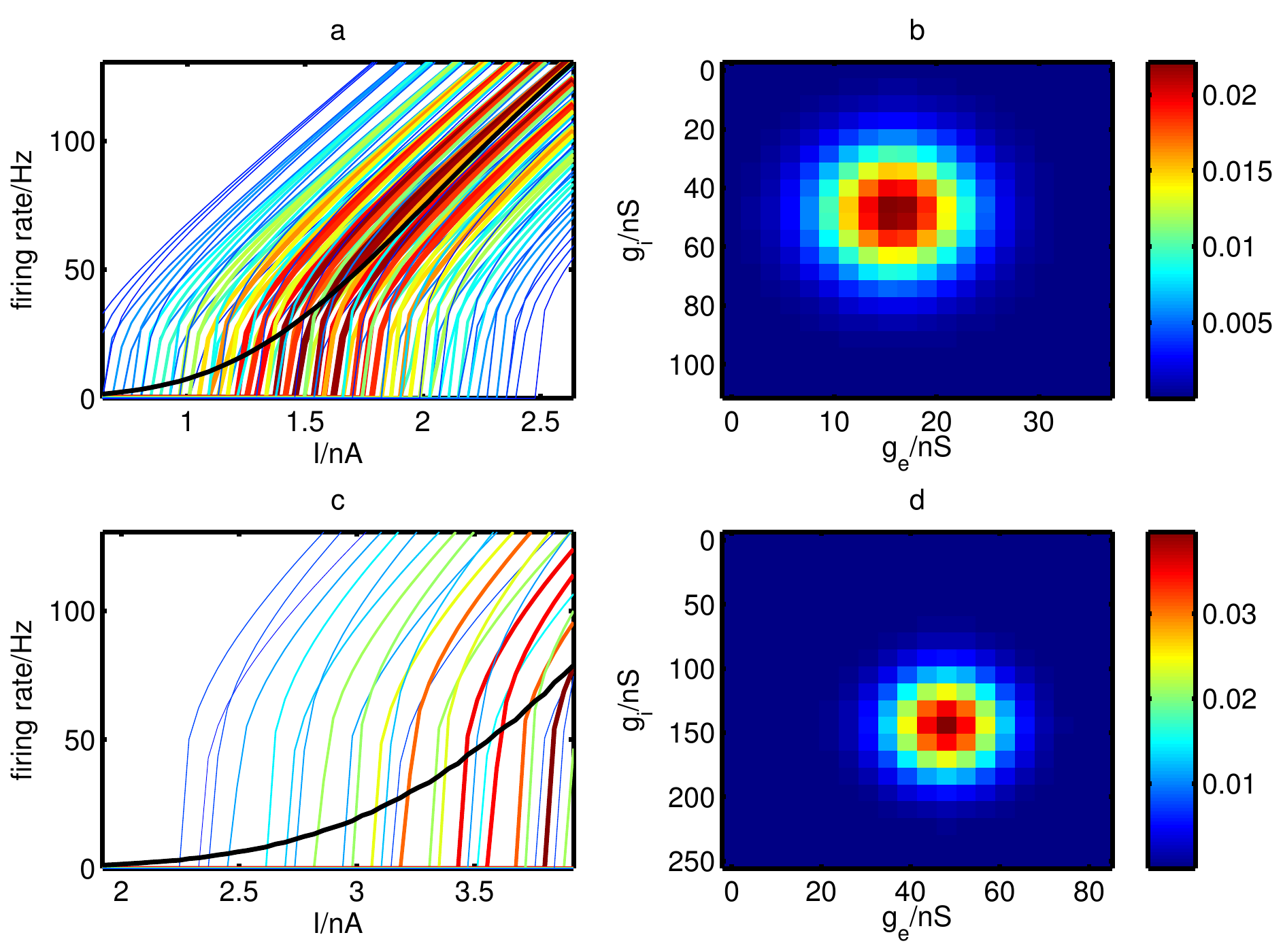}
	 \caption{\textbf{Illustration of how the weighted average of LIF firing rates is calculated to approximate the sLIF firing rate.}
	 \textbf{(a)} Deterministic LIF firing rates (\ref{eqn_f(I)}) for a range of conductance values $(g_e, g_i)$ at synaptic input rates $\lambda_e = \lambda_i = 1~\mathrm{kHz}$, colour and line thickness according to the joint probability distribution of the conductance values. The sampling density has been kept low for illustrative purposes. Black curve is the weighted average (\ref{eqn_weightedAverage}) of the colored curves, with the relative weight of each curve given by its colour (as in (b)) and (approximately) its line thickness. Only curves with relative weight $>e^{-2}$ are shown. Other parameters as in main text, section \ref{modelParameters}.
	 \textbf{(b)} Joint probability distribution (2D Gaussian) of $(g_e, g_i)$.
	 \textbf{(c,d)} As (a,b), but with $\lambda_e = \lambda_i = 3~\mathrm{kHz}$.
	 }
	 \label{figWeightedAverageIllustration}
	\end{figure}
	
	\begin{figure}[]
	 \includegraphics[scale=0.75]{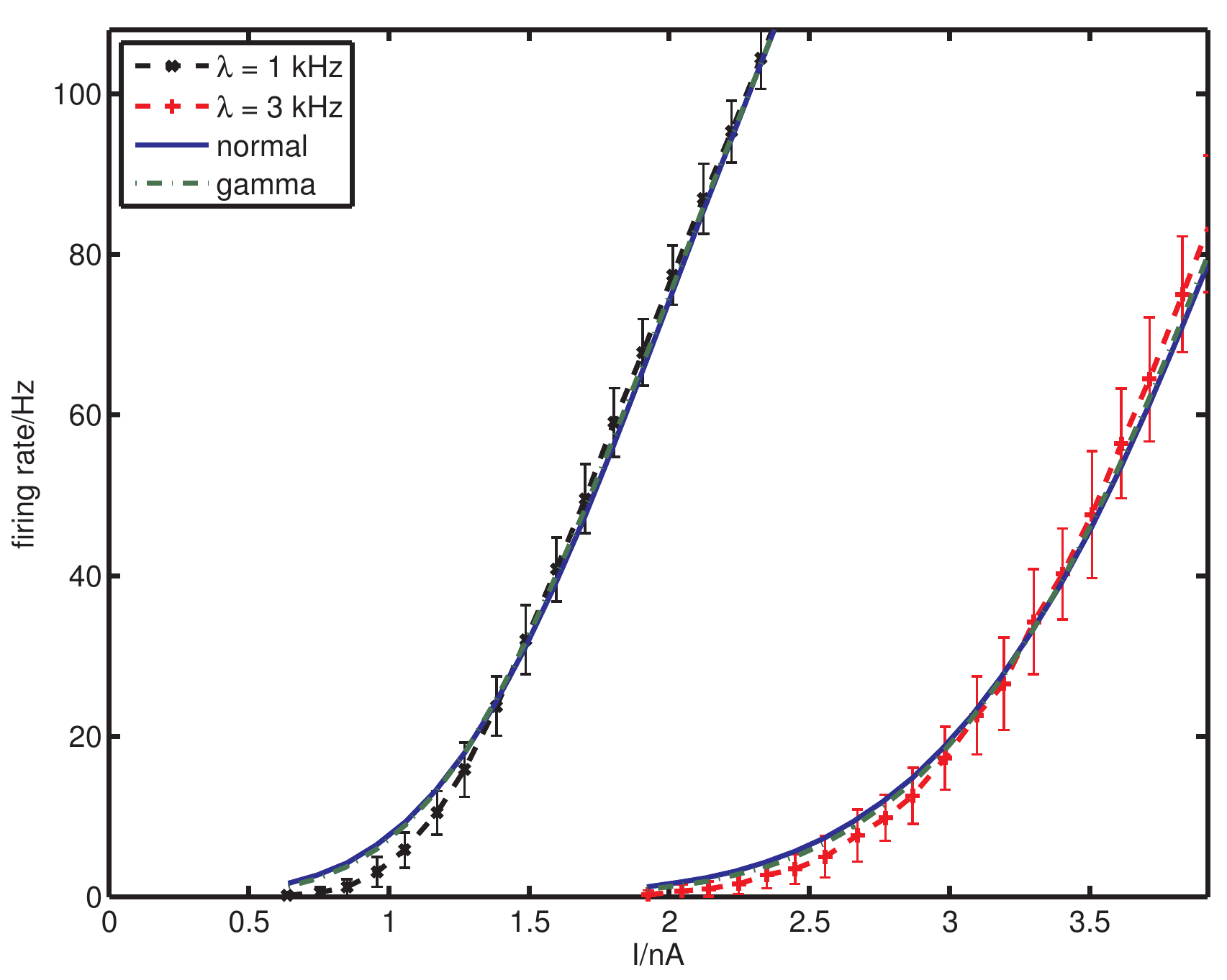}
	 \caption{\textbf{Firing rate of sLIF model simulations is approximated by weighted average of deterministic LIF firing rate expression}
	 Firing rates of the sLIF model were estimated from simulations with low ($\lambda_e = \lambda_i = 1~\mathrm{kHz}$) and high ($\lambda_e = \lambda_i = 3~\mathrm{kHz}$) noise balanced synaptic input (qualitatively similar fits could be produced for rates as low as $0.5~\mathrm{kHz}$). Error bars are standard deviation over 100 simulations. Simulation results are approximated by averaging the analytical firing rate expression for the deterministic LIF model over a normal or gamma distribution for the conductances.
	 }
	 \label{figWeightedAverageFiringRate}
	\end{figure}
	
	\subsection{Agreement of analytical approximation with simulations} 
		The steady state distribution of synaptic conductances under poissonian spike input is well approximated by the normal and gamma distributions derived in the previous section (Fig.~\ref{figConductanceDistributions}(a)). We can also approximate the distributions of $g$, $I$ and $g/I$ well (Fig.~\ref{figFurtherDistributions}(b-d)), but as discussed in section \ref{furtherDistributions} this is really only of use if we could then go on to derive the distribution of $\log(1-g(V_{th} - V_L)/I)$.
		
		Approximating $g_e$ and $g_i$ as either normal or gamma distributed, one can then numerically integrate (\ref{eqn_f(I)}) over the (separable) joint probability distribution $P(g_e,g_i)$ and thus in turn approximate the firing rate of a sLIF neuron to a given input, as illustrated in Fig.~\ref{figWeightedAverageIllustration}. The expression for the weighted average of the deterministic LIF neuron firing rate (\ref{eqn_weightedAverage}) agrees well with the firing rates estimated from direct simulation of the stochastic LIF equations and captures the decrease in gain when noisiness of the synaptic input is increased (Fig.~\ref{figWeightedAverageFiringRate}).

	\subsection{Relating parameters of the sigmoid non-linearity to those of the sLIF model} \label{similarityWithSigmoid}
		Having an analytical approximation for the firing rate of a sLIF neuron (\ref{eqn_weightedAverage}) that agrees well with the simulations (Fig.~\ref{figWeightedAverageFiringRate}) we would like to relate changes in the synaptic input statistics to the changes in the shape of the response curve. The weighted average curves resulting from (\ref{eqn_weightedAverage}) are similar in shape to the sigmoid non-linearities used to describe gain control in recent experimental work~\citep{Rabinowitz2012,Rabinowitz2011a}:
		\begin{equation}
			y(z) = a + \frac{b}{1 + e^{-\frac{z-c}{d}}}
			\label{eqn_sigmoid}
		\end{equation}
		where $a,b,c$ and $d$ are parameters.
		With this expression, gain control was studied by fitting the parameters of the sigmoid function as functions of stimulus contrast~\citep{Rabinowitz2012,Rabinowitz2011a} (Fig.~\ref{figGainKernelModel}).
		We would like to relate the parameters of this phenomenological description of gain control to parameters in our mechanistic model. Generally this is intractable, as (\ref{eqn_weightedAverage}) cannot be evaluated in closed form, but we can make some progress by comparing the shape of the weighted average firing rate with that of the sigmoid, and how their shapes change with respect to their parameters. In the sLIF model described here, there is no background firing, so $a = 0$, although this would have to be fitted to some positive value when comparing against \textsl{in vivo} data.
		
		Data from \textsl{in vivo} experiments mainly seem to lie on the left half of the fitted sigmoid function, i.e. the abscissae are smaller than the abscissa of the inflection point~\citep{Rabinowitz2012,Rabinowitz2011a}. By inspecting Fig.~\ref{figWeightedAverageIllustration} we see that the inflection abscissa of a sigmoid (were its left half fitted to the numerical data) roughly coincides with the threshold input to get firing (the rheobase) in the deterministic case, at mean conductance values. Intuitively, this can be explained by the fact that the deterministic firing rate curve for the mean conductances has the highest weighting in the average and hence strongly determines it shape. For abscissa greater than the mean threshold input, this main curve will be linear, i.e. with no curvature, as will the resulting average, approximately. Therefore we might be tempted to set $c$ to the threshold vale of our input $I$, i.e.
		\begin{equation}
			\mu_{th} = (\mu_{g_e} + \mu_{g_i} + g_L)(V_{th} - V_L)
			\label{eqn_c}
		\end{equation}
		Similarly one can argue that the standard deviation in the input threshold will be inversely related to the gain of the firing rate curve, because it controls the horizontal spread of deterministic firing rate curves with high weighting (Fig.~\ref{figWeightedAverageIllustration}), so one might like to try and set $d$ equal to
		\begin{equation}
			\sigma_{th} = \sqrt{\sigma_{g_e}^2 + \sigma_{g_i}^2}(V_{th} - V_L)
			\label{eqn_d}
		\end{equation}
		Similar observations about the qualitative dependence of the firing rate curve shape on input statistics have been made by \citet{Yu2003}, but they did not attempt to relate the f-I curve parameters to the input statistics in a functional form.
		
\begin{figure}[]
			\includegraphics[scale=0.75]{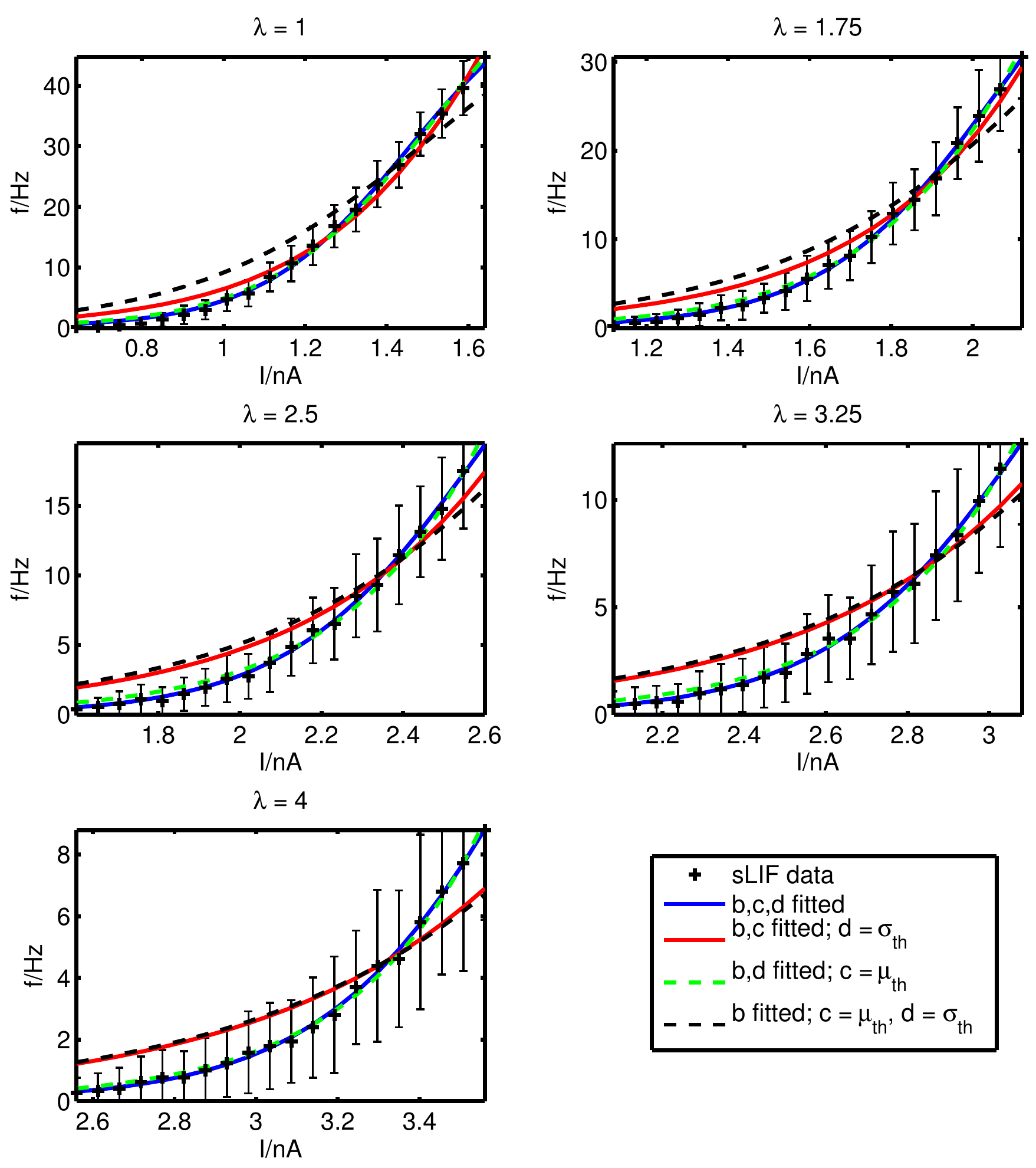}
			\caption{\textbf{sLIF firing rates are approximated by sigmoid functions} Data from simulations of the sLIF model for five balanced synaptic input rates ($\lambda = \lambda_e = \lambda_i$) were used to fit a sigmoid (\ref{eqn_sigmoid}) with $a = 0$ and the parameters $b, c ,d$ fitted, or with either or both $c$ and $d$ fixed at $\mu_{th}$ (\ref{eqn_c}) and $\sigma_{th}$ (\ref{eqn_d}), respectively. Error bars are standard deviations of 100 simulations.}
			\label{figCompareWithSigmoidFit}
		\end{figure}

\begin{figure}[]
			\includegraphics[scale=0.75]{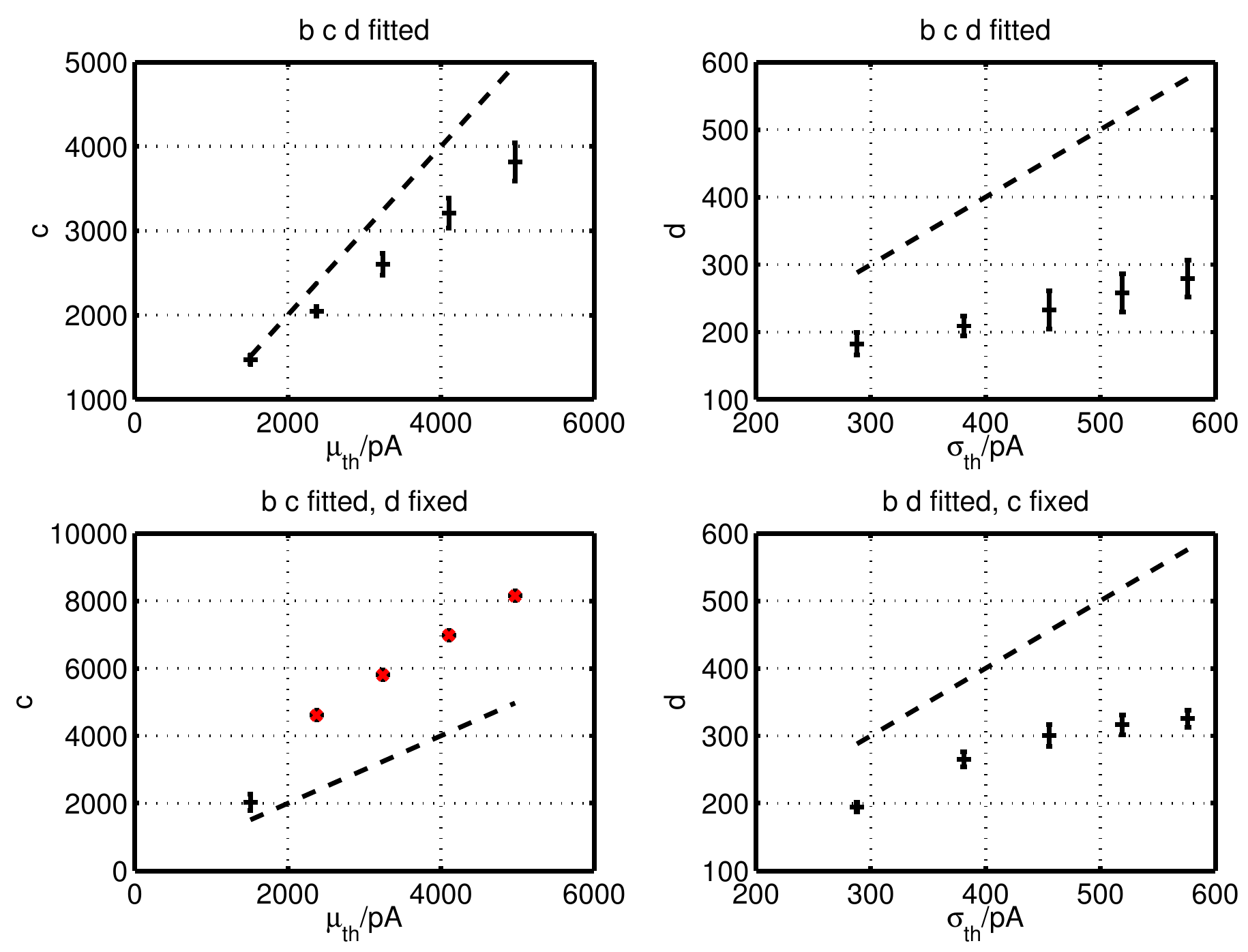}
			\caption{\textbf{Relationship between fitted sigmoid parameters and statistics of sLIF model parameters}
			  Parameters $b$, $c$ and $d$ of a sigmoid (\ref{eqn_sigmoid}) on data from sLIF model simulations for a range of balanced synaptic input rates from $\lambda = 1~\mathrm{kHz}$ to $4~\mathrm{kHz}$. Error bars are $95\%$ confidence intervals on fitted parameters, as given by \textsc{Matlab}'s \mcode{fit} function (curve fitting toolbox). Red diagonal crosses without error bars indicate that the confidence interval was greater than the range of abscissae. $\mu_{th} = (\mu_{g_e} + \mu_{g_i} + g_L)(V_{th} - V_L)$ and $\sigma_{th} = \sqrt{\sigma_{g_e}^2 + \sigma_{g_i}^2}(V_{th} - V_L)$.
			  }
			\label{figCompareWithSigmoidParameters}
		\end{figure}

		To test these predictions, we fitted sigmoid functions to data from sLIF simulations using \textsc{Matlab}'s \mcode{fit} function. The fits were generally qualitatively indistinguishable when initialised at random multiple times with $c$ and $d$ constrained to be of the same order of magnitude as $\mu_{th}$ and $\sigma_{th}$, though fitting would occasionally get stuck in what appeared to be local minima. For Figs~\ref{figCompareWithSigmoidFit}\&\ref{figCompareWithSigmoidParameters} $c$ and $d$ were initialised at $\mu_{th}$ and $\sigma_{th}$, respectively.

		A sigmoid with $c$ given by (\ref{eqn_c}) is in decent agreement with the sLIF neuron model data (Fig.~\ref{figCompareWithSigmoidFit}). When $d$ is given by (\ref{eqn_d}) agreement is less good but the qualitative characteristics as the curve shape and its shifting and scaling are still captured. The quality of fits is similar when both $c$ and $d$ are fixed and only parameter $b$ is fitted. 
	\clearpage
\section{Discussion}
	We have presented here a stochastic LIF neuron as a mechanistic model for contrast gain control in neural auditory processing. In the model, subtractive and divisive normalisation of the firing rate curve arise simply from the dynamics of noisy synaptic input. This in itself is no new observation, but it has, to our knowledge, not been made use of to understand auditory contrast gain control. Only recently has the tuning of contrast sensitivity \textsl{in vivo} been found to be similar to the spectrotemporal receptive field (STRF)~\citep{Rabinowitz2012,Rabinowitz2011a}. This is indicative of local computation and leaves open the possibility of contrast gain control occurring intrinsically by feed-forward mechanisms only, i.e. without recurrent feedback -- which is exactly the case in the sLIF model discussed here. Thus a neuron in the auditory cortex may not compute the stimulus contrast directly, but exhibit at least part of its gain control just through its biophysics, in response to a change in contrast in the input it receives from lower levels of auditory processing.\smallskip
	
	Others have drawn similar conclusions in a more general context, and with a slightly different interpretation: \citet{Yu2003} suggest that ``the change of effective stimulus threshold in various statistical stimulus environments is the key factor underlying variance or contrast gain control''. \citet{Hong2008} attribute a neuron's intrinsic gain control due to the fact that ``inputs with different statistics interact with the nonlinearity of the system in different ways''.

	\subsection{Predictions}
	  This hypothesis makes testable predictions on the relationship between mean and variance of the input to the auditory cortex -- they ought to be proportional, as in a Poisson process -- which is something that might be measurable in the inferior colliculus, which sits upstream of the cortex in the auditory pathway.\smallskip

	  In addition, effective membrane conductance and current distributions could be measured intracellularly and compared to our approximations (Fig.~\ref{figConductanceDistributions}). If mean  conductances and internal model parameter values (section \ref{modelParameters}) of a cell are known, one could examine their relationship to the parameters of the sigmoid nonlinearity (fitted to \textsl{in vivo} data) and their dependence on stimulus contrast, as in \eqref{eqn_c} or \eqref{eqn_d}. This could in turn motivate (or criticise) the particular choice of a sigmoid function for a neuron's response non-linearity, rather than a curve of similar shape but different functional form.\smallskip
	  
	  The best fits of sigmoids to spike rate data seem to be achieved for models were the divisive and subtractive parameters, $c$ and $d$ in \eqref{eqn_sigmoid}, have the same stimulus contrast dependence~\citep{Rabinowitz2011a}. This emerges naturally from a local mechanism of gain control, as the subtractive and divisive parameters are determined by the same synaptic input and hence have the same spectrotemporal tuning. Based on the arguments presented in section \ref{similarityWithSigmoid}, we proposed relationships for the sigmoid parameters $c \propto \mu_g$ and $d \propto \sigma_g$, which is in qualitative agreement with other work~\citep{Longtin2002,Ayaz2009,Ostojic2011}. If these relationships hold, the relationship between spectrotemporal contrast kernels and receptive fields ought to be $\kappa^{(c)}_{tf} \approx k_{tf}$ and  $\kappa^{(d)}_{tf} \approx |k_{tf}|$, rather than $\kappa^{(c)}_{tf} = \kappa^{(d)}_{tf} \approx |k_{tf}|$, as \citet{Rabinowitz2011a} suggested. Whether this can be confirmed upon re-inspection of previous results~\citep{Rabinowitz2011a} remains to be seen.
	  
	\subsection{Model criticism}
	 To achieve purely divisive normalisation without a subtractive shift of the firing rate curve, the sLIF model needs balanced excitatory and inhibitory input \citep{Longtin2002}. This might seem like physiologically unlikely fine-tuning, but balanced synaptic input has been observed experimentally and simple mechanisms for maintaining the balance have been proposed~\citep{Vogels2011}. Furthermore, purely divisive normalisation is rarely seen \textsl{in vivo}~\citep{Rabinowitz2011a}.\smallskip
	  
	  \citet{Hong2008} have described gain control as a diffusion of the noiseless f-I curve, which is similar in form to our weighted average \eqref{eqn_weightedAverage} for normal distributions. However, \citet{Hong2008} never see a decrease in LIF firing rate at higher noise levels, which we and others \citep{Ayaz2009} do observe in our LIF neurons. This disagreement might come down to the particular relationship of mean vs.\ variance in the input statistics, and the resulting subtractive vs.\ divisive scaling of the f-I curve. However, our aim was to understand and explain the functional form of gain control seen in the auditory cortex~\citep{Rabinowitz2011a,Rabinowitz2012}, which does exhibit decreased firing rate at increased noise levels.\smallskip

	  Fitting a sigmoid non-linearity with some derived, rather than fitted, parameters seems to give good agreement with model data (Fig.~\ref{figCompareWithSigmoidFit}). However, the suggested correspondences $c = \mu_{th}$ (\ref{eqn_c}) and $d = \sigma_{th}$ (\ref{eqn_d}) appear not to be exact (Fig.~\ref{figCompareWithSigmoidParameters}), though (nearly) linear. Thus we might have to (a) rethink the argument presented in section \ref{similarityWithSigmoid} and consider the possibility that $c$ and $d$ are determined by the statistics of a different set of model parameters (which might well have a similar dependence on the synaptic input rates, but with different constants of proportionality) or (b) make the fitting of the sigmoid parameters more robust.

	\subsection{Impact beyond auditory neuroscience}
	  Beyond the application to auditory contrast gain control, we have derived an analytical approximation (\ref{eqn_f(I)}) to the firing rate of a stochastic LIF neuron that, if still not in closed form, is easy to evaluate (involving numerical integration over only two dimensions) and hopefully lends itself more easily to an intuitive understanding of how the input statistics affect the firing rate than other expressions~\citep{Ermentrout2010,Brunel1998,Lanska1994,Lansky1995,Longtin2002,Buonocore2010}.

	\subsection{Further work}
	  Promising directions for further work exist, most notably to fit the sLIF model on stimulus data that have been filtered through estimated STRFs (as in~\citep{Rabinowitz2012,Rabinowitz2011a}) and to refit the gain kernel model on data generated from the sLIF model. Mathematically, one could investigate whether membrane voltage distributions can be found when its evolution is described by a stochastic process with fluctuations of voltage-dependent magnitude.

	\subsection{Acknowledgements}
	LJS would like to thank Neil Rabinowitz, Ben Willmore and Jan Schnupp for their collaboration and helpful discussions.
	
\bibliographystyle{revtex4-1}
\bibliography{Neuro-ContrastGainControl}

\appendix
\section{Other approximations for the conductance distribution and beyond}
		\subsection{Gamma distribution}
			As seen in the results section (Fig.~\ref{figConductanceDistributions}), the steady state distribution of the conductances seems to be captured even better by a gamma distribution with shape $k$ and scale $\theta$, especially with respect to its skew. Matching the mean ($k\theta$) and variance ($k\theta^2$) of the gamma distribution to that of the normal approximation (\ref{eqn_mu_g},\ref{eqn_sigma_g}), we require the shape parameter to be $k = 2\lambda\tau_g$ and the scale $\theta = \Delta g/2$. It is tempting to try to arrive at these parameters by deriving the gamma distribution as the probability distribution of the sum of $k$ exponentially distributed random variables of mean $\theta$, or as the distribution of waiting times until `death' of random variables with poissonian arrival. However, the analogy does not seem to work, the parameters would be out by factors of 2 or 1/2.
			For this reason approximating the steady-state conductances as gamma distributed it less appealing. To arrive at an expression for the distribution of firing rates we would also like to know, as a second step, the distribution of sums and differences of the conductances (as these appear as $g$ and $I$ in the firing rate expression (\ref{eqn_f(I)})), but there seems to be no established result for the difference of two gamma distributed random variables, and the sum for different shape parameters is an unwieldy expression~\citep{Coelho1998}.
			
		\subsection{Further distributions} \label{furtherDistributions}
\begin{figure}[] 
	 \includegraphics[scale=0.75]{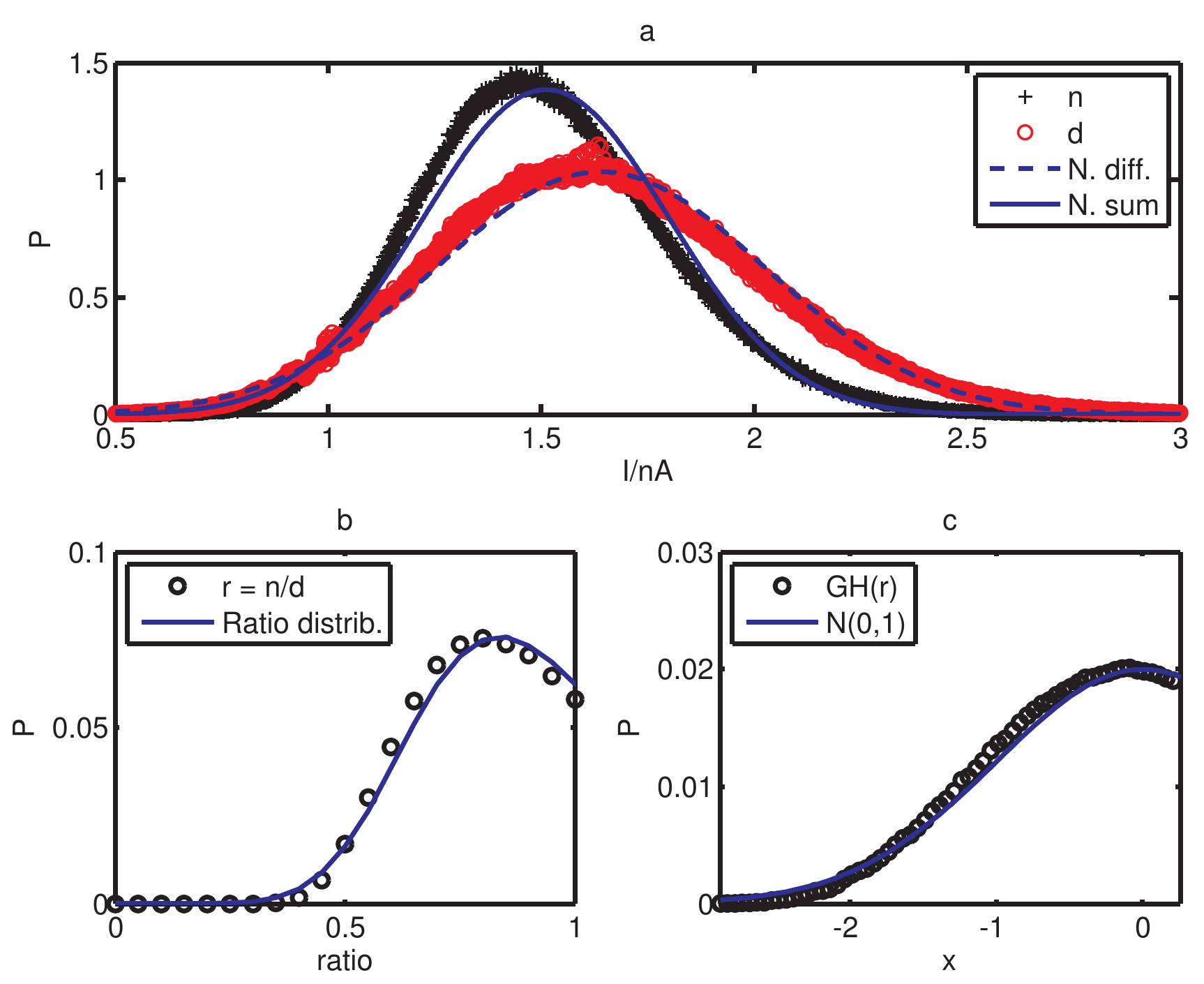} 
	 \caption{\textbf{Steady-state distributions of terms involving the excitatory ($g_e$) and inhibitory ($g_i$) conductances in a sLIF neuron model.} 			The model differential equation, in terms of the parameters plotted, is $C\dbyd{V}{t} = d - \frac{n}{V_{th} - V_L}V$ and the firing rate for the deterministic case is $f(I) = {[} t_{ref} - \frac{C(V_{th} - V_L)}{n(g)}\log(1 - r(I,g)) {]}^{-1}$. Distribution parameters are \textsl{not} fitted, but derived as described in the main text (\ref{eqn_mu_g},\ref{eqn_sigma_g}). Data shown for synaptic input rates $\lambda_e = \lambda_i = 1~\mathrm{kHz}$, but approximations were qualitatively similar for rates as low as $0.5~\mathrm{kHz}$, and better the higher the rates.
	 	\textbf{(a)} Effective current $d = I_{FF} + I_e - I_i$ and effective conductance (multiplied by a voltage term) $n  =  g(V_{th} - V_L)$ are approximately normally distributed, with $\mu_d \propto \mu_{I_e} - \mu_{I_i}$, $\sigma_d^2 = \sigma_{I_e}^2 + \sigma_{I_i}^2$ and $\mu_n \propto \mu_{g_e} + \mu_{g_i}$, $\sigma_n^2 = \sigma_{g_e}^2 + \sigma_{g_i}^2$
		\textbf{(b)} The distribution of the ratio $r$ of the terms in (b) approximately follows the ratio distribution~\citep{Hinkley1969}. Range shown is limited to $f(I)>0$.
		\textbf{(c)} The Geary-Hinkley-transform~\citep{Geary1930,Hinkley1969} makes $r$ approximately normally distributed with zero mean and unit variance. Maximum abscissa as in (c).
		}
	 \label{figFurtherDistributions}
	\end{figure}

			Approximating the distribution of $g_e$ and $g_i$ as normal, we can work out the distribution of $g$ and $I$, which will also be normal and (nearly) independent. The next term in the firing rate expression (\ref{eqn_f(I)}) whose distribution we would like to know is $g(t)(V_{th} - V_L)/I(t)$. This is approximated by a Gaussian ratio distribution~\citep{Hinkley1969}, which can be made me made more Gaussian by a Geary-Hinkley-transformation~\citep{Geary1930,Hinkley1969} of the random variable as long as $I(t)$ is unlikely to be negative. As seen in Fig.~\ref{figFurtherDistributions}, this all works well, but going further is difficult. Working out the distribution of $\log(1 - g(t)(V_{th} - V_L)/I(t))$ while only considering values for which $g(t)(V_{th} - V_L)/I(t) < 1$ is tricky and goes beyond the scope of this paper. Instead, we turn to a more pragmatic approach in the next section.

\end{document}